\documentclass[aip,apm,amsmath,amssymb,twocolumn,reprint]{revtex4-1}
\usepackage{graphicx}% Include figure files
\usepackage{bm}% bold math
\usepackage{amsmath}
\usepackage{physics}
\usepackage{mathrsfs}
\usepackage{calrsfs}
\begin{document}

\title{On the origin of near infrared photoluminescence in Bismuth-doped aluminosilicate glass}

\author{Oleksii Laguta}
\affiliation{CERLA, PHLAM UMR CNRS 8523, University Lille-1, Lille 59655, France}
\author{Boris~I.~Denker}
\affiliation{A. M. Prokhorov General Physics Institute of RAS, Vavilov str. 38, Moscow 119991, Russia}
\author{Boris~I.~Galagan}
\affiliation{A. M. Prokhorov General Physics Institute of RAS, Vavilov str. 38, Moscow 119991, Russia}
\author{Sergei~E.~Sverchkov}
\affiliation{A. M. Prokhorov General Physics Institute of RAS, Vavilov str. 38, Moscow 119991, Russia}
\author{Igor~M.~Razdobreev}
\email{Igor.Razdobreev@univ-lille1.fr}
\affiliation{CERLA, PHLAM UMR CNRS 8523, University Lille-1, Lille 59655, France}

\date{Compiled \today}

\begin{abstract}
	The near infrared (NIR) photoluminescence (PL) in Bismuth-doped aluminosilicate glasses was reported about of 16 years ago. 
	Although it has since been developed various kinds of fiber lasers and amplifiers based on the Bismuth-doped silica glasses, 
	the main problem, namely the nature of NIR PL, remains unsolved. Here we show that this famous NIR PL in Bismuth-doped 
	aluminosilicate glass originates from the defect, most probably positively charged oxygen vacancy V$_{O}^{2+}$ in the glass 
	network. This defect interacts with Bi$^+$ ion in its environment via the energy transfer. 
	
\end{abstract}
	
	\maketitle
	
\noindent Interest in Bismuth-doped materials is caused by a multitude of potential applications. Perhaps the most exciting application 
of Bi-doped materials is in the area of fiber lasers (BFL) and amplifiers (BFA) that cover inaccessible by the rare earths spectral 
windows \cite{Bufetov2009}. Nevertheless, such devices suffer from a number of drawbacks. The very low levels of Bismuth 
doping and, as a consequence, significant fiber length (typically 80\,-\,100\,m) are necessary to ensure the efficient BFL and 
BFA operation. Also, the efficiency of the BFL and BFA remains significantly lower in comparison to their rare earth counterparts. 
Poor understanding of the nature of luminescent centers in Bi-doped  glasses does not allow the development of efficient devices. 
Unfortunately, since the first demonstration of the near infrared (NIR) photoluminescence (PL) in Bi-doped silica glasses 
\cite{Fujimoto2001} and up to now its nature remains a subject of controversies \cite{Peng2011a,Dianov2012a}.

Recently \cite{Laguta2015b}, some of us investigated the magnetic circular polarization of luminescence (MCPL) in Bi-doped pure 
silica glass with low Bismuth content and without other dopant. It was shown that the lasing excited state (ES) is a magnetic multiplet 
and the temperature dependence of the degree of MCPL was in excellent agreement with the model of  emission from an isolated
non-Kramers doublet. The latter implies that the center responsible for the lasing is an even electron system. Furthermore, the spectral
features of MCPL allow to assume that the ground state (GS) is also magnetic multiplet that in turn denies Bi$^+$ ions as a possible 
origin of NIR PL. Unfortunately,  the extremely low content of Bismuth in studied samples did not allow to probe the active centers with 
magnetic circular dichroism (MCD) technique. In contrast to MCPL, MCD technique \cite{Stephens1974a,Riehl1976a} probes directly 
GS of an active center and it could be able to solve definitely this problem. Very recently \cite{Laguta2015c} we also shown with both 
MCPL and MCD techniques, that in heavily doped \textit{aluminosilicate} glass there exist three centers, and all of which can be 
described as even-electron center. In the present Letter we report the investigation of Bi-doped aluminosilicate system, namely, 
MgO-Al$_2$O$_3$-SiO$_2$ glass \cite{Denker2009}, with a relatively high  content of Bismuth. For the first time our experiments 
unambiguously prove that in \mbox{Bi-doped} \textit{aluminosilicate} glass there is a system of two interacting centers which 
consists of Bi$^+$ ion and some defect in the glass network, the latter being induced or modified by the Bismuth doping. This defect 
is responsible for the strong absorption in the visible range of spectrum (maximum at 500\,nm) and intense luminescence in the NIR 
which was the subject of long debates. Finally, the third center, clearly detected in MCD spectrum, can be put in correspondence to 
clusters of Bismuth ions. 

The glass synthesis was performed in corundum crucibles at 1470 $^{\circ}$C following the procedure described previously
\cite{Denker2009}.  In the present study the following basic composition of aluminosilicate glass was chosen \mbox{22.2 wt.\%} MgO +
\mbox{22.2 wt.\%} Al$_2$O$_3$ and \mbox{55.6 wt.\%} SiO$_2$. To this content 3, 1, 0.5, 0.125 or 0 mol.\% of Bi$_2$O$_3$ 
were added. All results reported in this Letter were obtained with sample containing 0.5 mol.\% of Bi$_2$O$_3$. 

\begin{figure}[htbp]
	\centering
	\includegraphics[width=8.4cm]{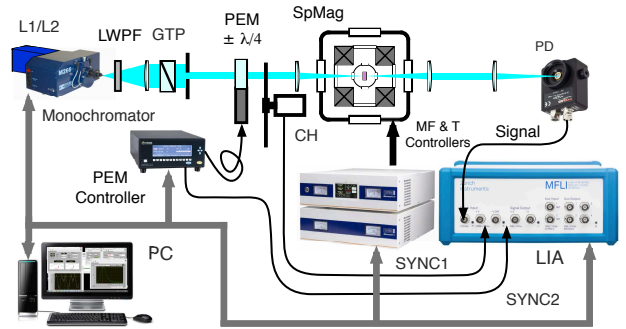}
	\caption{MCD experimental setup. L1/L2 - UV Xenon or quartz tungsten lamp, LWPF - long wave pass filter; 
		GTP - Glan-Thompson polarizer; PEM - photo-elastic modulator;  CH - chopper; SpMag - cryomagnetic system; 
		PD - Si or Ge amplified photo-detector; MFLI - multi-frequency lock-in amplifier.}
	\label{Setup}
\end{figure}

The experimental setup for MCD measurements is shown schematically in Fig.\,\ref{Setup}. The experiments 
in the temperature range of 1.4\,-\,280\,K and magnetic fields (MF) in the range of  0\,-\,6.5\,T were performed in the closed 
cycle magneto-optical cryostat (SpectromagPT, Oxford Instr.). The thermal stability of the samples attached 
to the holder of the variable temperature insert  was about of 0.01\,K except the range from 4.2 to 10\,K where  
the thermal stability was $\sim$\,0.05\,K.  The quartz tungsten or UV enhanced Xenon lamps were used as a 
source filtered with a single grating monochromator. After the exchangeable long wave pass filter (LWPF) and 
collimating lens the light beam was polarized with a large aperture Glan-Thompson polarizer (GTP). The central
part of the beam selected by diaphragm was modulated first by the photo-elastic modulator (PEM)  at 
quarter-wave peak retardation at the frequency 20.077\,kHz (I/FS-20, Hinds Instr.) then chopped at 400 Hz. 
The beam impinges on the sample located in a magneto-optical cryostat,  then focused on the amplified silicon 
or Ge photodiode (PDA36A or PDA30B, Thorlabs). Two resulting signals at 400 and 20\,kHz were demodulated 
simultaneously by the multi-frequency lock-in amplifier (MFLI, Zurich Instr.). The spectral resolution in the range 
320\,-\,900\,nm  was $\sim$ 2.2\,nm, and 5 nm between 900 and 1550\,nm. The signal of MCD measured in
our experiments was defined as  $\Delta$A = V$_{20kHz} $/ V$_{400Hz}$, where $ V_{20kHz}$ and $V_{400Hz}$ 
are the signals obtained directly from the lock-in amplifier at corresponding frequency \cite{Mason2007}. 
The excitation and PL spectra were measured following the procedure described previously \cite{Laguta2015a}.

\begin{figure}[htb]
	\centering
	\includegraphics[width=8.4cm]{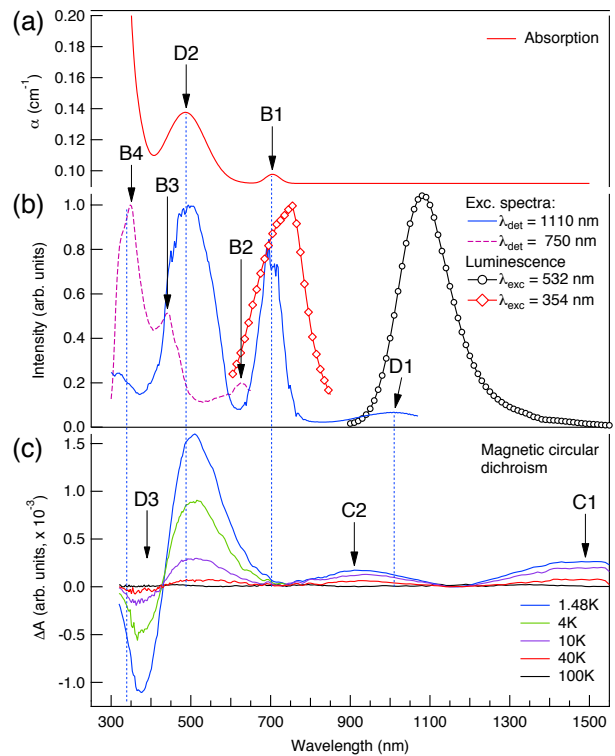}
	\caption{a) Absorption spectrum; b) Normalized excitation and PL spectra at T\,=\,300\,K.  
		c) MCD spectra at fixed MF of 6\,T. Each series of peaks is named as follows: Bi for Bi$^+$ ion, Di for \lq\lq defect\rq\rq and Ci for  \lq\lq clusters\rq\rq (see text).
	}
	\label{ABS}
\end{figure}

In Fig.\,\ref{ABS}(a) and (b) we show the absorption and excitation spectra, respectively, 
measured at room temperature. It is seen that only two absorption bands with maximums at 488 (D2) and 703\,nm
(B1) can be detected in the range 400\,-\,1550\,nm. These absorption bands coincide with two bands in the excitation 
spectrum recorded at 1110\,nm that corresponds to the NIR PL band shown in Fig.\,\ref{ABS}(b).
The excitation spectrum reveals an additional band in the NIR spectral region with the maximum at 1030\,nm, 
that corresponds to the Stokes shift for the NIR PL band \mbox{$\approx$ 700\,cm$^{-1}$}.
At first glance both D2 and B1 bands belong to the same NIR PL center. 
Nevertheless, excitation spectrum recorded at 750\,nm corresponding to the "red" PL band, also shown in 
Fig.\,\ref{ABS}(b), reveals three bands at 626 (B2), 441 (B3) and 339\,nm (B4) that do not coincide with 
NIR PL excitation spectrum. This discrepancy can be explained by the presence of two centers with the 
energy transfer between them. This assumption is supported by the MCD spectra shown in Fig.\,\ref{ABS}(c).
First, all the observed bands in the spectrum are temperature dependent that implies magnetic multiplets in 
the corresponding GS's. Second, only the strongest  MCD band at 510\,nm can be put in direct 
correspondence to the absorption and excitation band D2. Third,  there is no MCD bands that could correlate with 
one of the bands in B series. It follows immediately that D2 band corresponds to the center with the magnetic multiplet 
in its GS. Thus, band D2 cannot be assigned to Bi$^+$ ion as it has been proposed in many anterior works, for example 
in \cite{Zhou2008a}. On the contrary, band B1 can be assigned to the E state ($^3$P$_1$, M$_J$=$\pm$1 in C$_{3v}$ 
local symmetry) of Bi$^+$ ion because its GS is a magnetic singlet and we do not expect to observe the effect of MCD 
for this ion.  Also, this transition can be sufficiently strong \cite{Davis1967,Bartram1989b}. Moreover, it has been shown 
in \cite{Razdobreev2014a} that in Ga/Bi co-doped silica glass, which is a counterpart of the studied here Bismuth-doped 
aluminosilicate glass, the main relaxation channel of the state corresponding to the band B1 is the energy transfer due 
to the strong quadrupole-quadrupole interaction with another center. This explains the appearance of band B1 in the excitation 
spectrum recorded at 1110\,nm. Returning to Fig.\,\ref{ABS}(c), one can see that MCD spectrum reveal the additional
bands:  at 374\,nm (D3), 910\,nm (C2), and a very large band (C1) from 1200\,nm up to the limit of available spectral range. 
This  latter band has a flat maximum between 1420 and 1520\,nm and, most probably, it consists of at least two overlapping 
bands. The bands C1 and C2 cannot be assigned to one of the above series (centers). The details on the origin of 
MCD bands were obtained investigating their  temperature and MF dependences.

\begin{figure*}[htb]
	\centering
	\includegraphics[width=10cm]{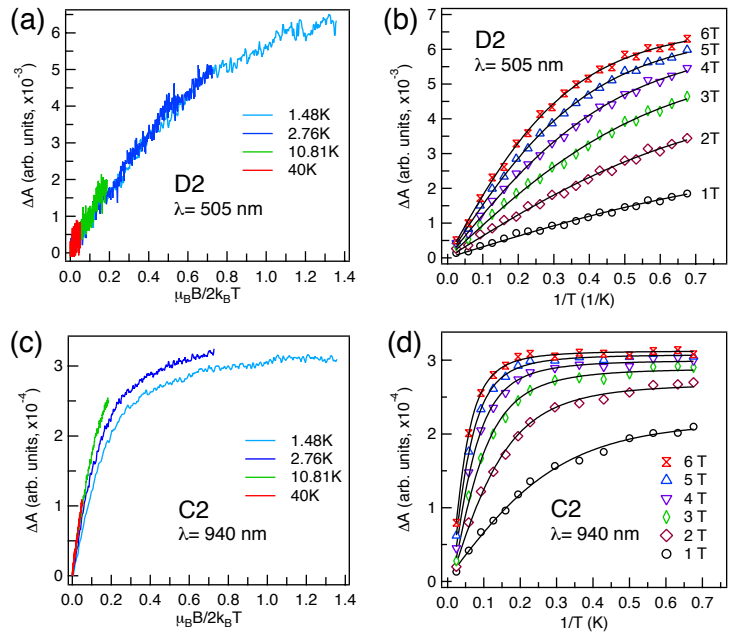}
	\caption{Temperature and magnetic effects on the GS's of optical centers.
		The saturation magnetization curves in a) and c) are shown for a variable MF and fixed temperatures. 
		(b) and (d) are the replot of data on the left in terms of temperature dependence at constant field.
		Solid lines in (b) and (d) are the theoretical profiles obtained with the global fit to the orientation averaging integral (\ref{eq}).}
	\label{MCD}
\end{figure*}

In Fig.\,\ref{MCD} we show the MF dependences of $\Delta$A plotted as a function 
of $\mu_{B}B/2k_{B}T$ at  fixed temperatures, where $\mu_{B}$ is the Bohr magneton, and $k_{B}$ is 
the Boltzmann constant. It is seen that for MCD band D2 at 505\,nm the isotherms corresponding to various temperatures 
are nearly coincident (Fig.\,\ref{MCD}(a)). On the contrary, for NIR MCD band C2 the isotherms exhibit pronounced nesting 
behavior and saturation at low temperatures (Fig.\,\ref{MCD}(c)). In Fig.\,\ref{MCD}(b) and (d)  the corresponding data from 
the left panels of the figure were replotted to show the saturation data in terms of temperature dependence at 
constant field. It is seen that for the band D2 the magnetization does not attain saturation at any field. In contrast,
the magnetization easily saturates in the NIR band C2 resulting in distinct  values of $\Delta$A. The behavior of bands D3 and 
C1 was very close to that of D2 and C2, respectively. For this reason we assigned  D2 and D3 to the same series, while 
C1 and C2 belongs to another one. 
The behavior exhibited by the MCD bands D2, D3, C1, C2 can be explained with the assumption that both centers (D and C)
are the even-electron systems (see more detailed discussion in \cite{Laguta2015b}). The data in Fig.\,\ref{MCD}(b) and (d)
were fitted together with that for D3 and C1 MCD bands (not shown in  Fig.\,\ref{MCD}) to the following orientation averaging 
integral \cite{Solomon1995a}:
\begin{equation}
\Delta A \\ 
=A_{sat}\int_0^1\!\frac{2n^{2}G_z}{\sqrt{\Delta^{2}+4n^{2}G_z^{2}}}\tanh\left( \frac{\sqrt{\Delta^{2}+4n^{2}G_z^{2}}}{2k_{B}T}\right)\mathrm{d}n,  
\label{eq}
\end{equation}
\noindent where $G_z  = g_{\parallel}\mu_{B}B$, $n =\cos\theta$, $\Delta$ is the splitting between two magnetic sub-levels due 
to the rhombic component of the crystal field and 
A$_{sat}$ is the saturation constant. The data fit results are reported in Table\,\ref{tab} and the corresponding profiles of $\Delta$A 
as a function of inverse temperature are shown in Fig.\,\ref{MCD}. It is seen that MCD bands Ci and Di belong to different centers, 
as already has been noted above. Given a sufficiently large value of g-factor obtained for Ci bands, we believe that they should be 
assigned to clusters of Bismuth ions. It is also worth noting that all the values of zero-field splitting (ZFS) are too large to observe
the electron spin resonance (ESR) in the most commonly used X- or even Q-band. 
\begin{table}[htbp]
	\centering
	\caption{\bf Zero-field splittings and $g$-factors in ground states}
	\begin{tabular}{lccc}
		\hline
		Center & Wavelength & $\Delta$\ & $g$  \\
		& (nm) &(cm$^{-1}$) &   \\
		\hline
		%		D (MCPL)           & 1110         & ES  & 43.5 $\pm$ 1.7     & 1 $\pm$ 3.6  \\
		D (defect)         & 375, 505   & 1.36 $\pm$ 0.13   & 0.9 $\pm$ 0.03  \\
		%		B (MCPL)        & 710           & ES  & 8.45 $\pm$ 0.01   & 1.29 $\pm$10$^{-5}$ \\
		C (cluster)          & 940, 1500 &  2.77 $\pm$ 0.02  & 4.4 $\pm$ 0.02 \\
		\hline
	\end{tabular}
	\label{tab}
\end{table}

\begin{figure}[htb]
	\centering
	\includegraphics[width=8cm]{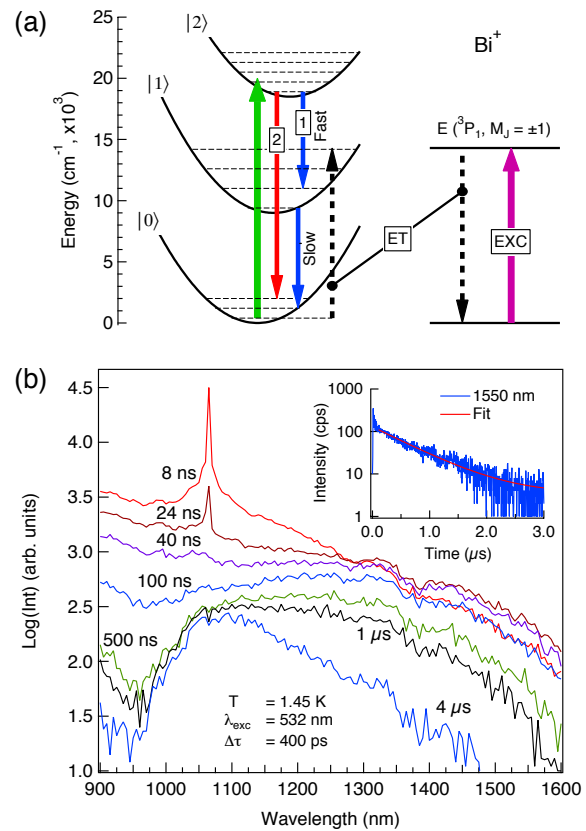}
	\caption{a) Energy level diagram of NIR luminescent center. Also shown the energy transfer (ET),
		which follows Bi$^+$ resonant excitation (EXC). b) Time resolved spectra under pulsed excitation. Inset: fast kinetics
		at 1550 nm.}
	\label{Diagram}
\end{figure}

We assign all the Di peaks in the excitation and MCD spectra, as well as the NIR PL at 1110\,nm, to the defect in the 
glass network. In Fig.\,\ref{Diagram}(a) we show schematically the energy levels of this defect interacting with Bi$^+$ ion 
in its environment. Here we show only two ES's of defect and the first ES multiplet of Bi$^+$ ion. It is clear from the absorption  
spectrum shown in Fig.\,\ref{ABS}(a) that the transition $\ket{0}\!\leftrightarrow\!\ket{2}$ is allowed. At the same time the transition
$\ket{0}\!\leftrightarrow\!\ket{1}$ is forbidden, because it does not appear neither in absorption nor in MCD spectra and the lifetime
of the NIR PL from the state $\ket{1}$ is very long. On the contrary, we can expect that $\ket{2}\!\leftrightarrow\!\ket{1}$ transition 
is allowed because of the very strong ES absorption from the state $\ket{1}$ for the wavelengths shorter than 1000\,nm \cite{Kalita2008a}. 
Thus, it is possible to observe two fast (allowed) transitions in the emission: $\ket{2}\!\rightarrow\!\ket{1}$ and  
$\ket{2}\!\rightarrow\!\ket{0}$, shown with blue (1) and red (2) arrows in Fig.\,\ref{Diagram}(a). The first transition should 
span only NIR wavelength region, while the second one can be observed also in the visible. In Fig.\,\ref{Diagram}(b) we show 
the time resolved spectra recorded under sub-nanosecond pulse excitation (micro-chip single longitudinal laser with 
$\Delta\nu\!\simeq$1\,GHz) in the wavelength range 900\,-\,1600\,nm. At the initial stage of PL, due to the intense monochromatic 
excitation, one can observe the fluorescence line narrowing in the spectrum, which is dominated by two fast components. The short 
and long 
cut-off wavelengths for the forbidden transition $\ket{1}\!\rightarrow\!\ket{0}$ are $\sim$\,950 and 1500\,nm, respectively. 
The PL recorded between 1500 and 1600\,nm corresponds to the transition $\ket{2}\!\rightarrow\!\ket{1}$ and its kinetics 
at 1550\,nm is shown in the inset. The single exponential fit resulted in the lifetime as short as 0.64\,$\pm$\,0.01\,$\mu$s.  
Transition $\ket{2}\!\rightarrow\!\ket{0}$ appears as a red luminescence with a trailing NIR wing under excitation in the state 
$\ket{2}$ of the defect. When excited at 354\,nm, this luminescence appears  as a short-wavelength shoulder of the \lq\lq red\rq\rq PL 
band shown in Fig.\,\ref{ABS}(b). The experimental proof of this statement, which follows from the analysis of MCPL 
is out of the scope of this Letter and will be reported elsewhere.

Point defects or localized states caused by the presence of 6$p$ (Bi, Pb) and 5$p$ (Sn, Sb) ions were considered 
for the first time by Sharonov et al. \cite{Sharonov2008,Sharonov2009} without emphasis on a particular nature of these
centers. They also pointed out that four major bands on excitation-emission plots do not necessarily belong to the same 
optical center. MCD experiments reported here directly confirm this assumption for aluminosilicate glass. Recently, 
Dianov et al. \cite{Dianov2015a,Sokolov2015a} suggested one more microscopical model of NIR PL centers in 
Bismuth-doped oxide glasses in which Bi$^{2+}$ and oxygen vacancy form a single molecular unit. In \cite{Laguta2015b} 
some of us have shown that this model fails to explain the NIR PL in Bismuth-doped pure silica. Presented here experiments show 
that it fails also to explain the NIR PL in Bismuth-doped aluminosilicate glass. First, the system is even electron and, 
second, Bi$^+$ ion and defect do not form a single center (molecule) preserving their individual properties. In our opinion,
the most probable route of the formation of optically active defect is the transfer of two electrons from the oxygen vacancy 
to Bi$^{3+}$ ion, thus forming a positively charged oxygen vacancy V$_{O}^{2+}$, responsible for the NIR PL, and Bi$^+$
ion in its environment. This can explain the low values of \mbox{$g$-factors} in GS and ES's of the NIR active center. 

In conclusion,  MCD was investigated in Bismuth-doped MgO-Al$_2$O$_3$-SiO$_2$ glass for the first time.
It was shown that Bismuth-doped  aluminosilicate glass contains at least three distinct centers. Two centers  were identified 
as  Bi$^+$ ion and associated defect that exchange their excitation via ET. The experiments indicate that the defect is an 
even-electron (or rather even-hole) system with magnetic multiplet in its GS and first ES, most probably positively charged 
oxygen vacancy V$_{O}^{2+}$ in the glass network . Finally, we identify the third center, which was observed only in MCD 
spectra, as clusters of Bismuth ions that lead to the quenching of lasing at increasing the concentration of Bismuth.

\section*{Funding}

O.V.L. and I.M.R. were supported by \lq\lq Agence Nationale de la Recherche\rq\rq~via ANR~grant  \lq\lq BOATS\rq\rq~12BS04-0019-01, 
and \lq\lq Fonds Europ\'een de D\'eveloppement Economique des R\'egions\rq\rq (FEDER) through the \lq\lq Contrat de 
	Projets Etat R\'egion (CPER) 2007-2013\rq\rq.\\ 
B.I.D., B.I.G. and S.E.S. were supported by RFBR (grant 14-02-00296-a) and Russian Academy of Sciences (Presidium program P1).

\end{document}